# Malware Detection in IOT Systems Using Machine Learning Techniques

Ali Mehrban[1] and Pegah Ahadian[2]

[1]School of Electrical and Electronic Engineering, Newcastle University, Newcastle, UK
a.mehrban@ieee.org
[2]Department of Computer Science, Kent State University, Ohio, USA
pahadian@kent.edu

*Malware detection in IoT environments necessitates robust methodologies. This study introduces a CNN-LSTM hybrid model for IoT malware identification and evaluates its performance against established methods. Leveraging K-fold cross-validation, the proposed approach achieved 95.5% accuracy, surpassing existing methods. The CNN algorithm enabled superior learning model construction, and the LSTM classifier exhibited heightened accuracy in classification. Comparative analysis against prevalent techniques demonstrated the efficacy of the proposed model, highlighting its potential for enhancing IoT security. The study advocates for future exploration of SVMs as alternatives, emphasizes the need for distributed detection strategies, and underscores the importance of predictive analyses for a more powerful IOT security. This research serves as a platform for developing more resilient security measures in IoT ecosystems.*

## 1. Introduction

The Internet of Things (IoT) is a powerful digital technology that connects the real and online worlds. It lets people, things, and machines communicate and interact with each other across the globe. This creates new ways of doing business and working together. But IoT devices are also hard to design and secure, so cybercriminals can easily hack them. They use weak passwords, old software, and malware to take over IoT devices [1]. In 2020, one out of every four cyberattacks was aimed at IoT devices, and this number will only go up as more people use these technologies. Malware is one of the biggest dangers for IoT devices, and it shows how important it is to have better security solutions.

For example, in October 2016, Dyn, a major DNS provider in the US, was hit by one of the biggest and strongest DDoS attacks by the Mirai malware family. This malware infected more than 1.2 million IoT devices, and attacked popular websites like Google and Amazon.

So, it is very important for researchers to improve the security of IoT devices, especially when they deal with IoT-related malware. There are many research studies on how to make IoT devices more secure, such as how to protect IoT communications [2][3]. Jamal Adineh and his colleagues have categorized various applications of the Internet of Things to identify security requirements and their upcoming challenges. They analyze traditional encryption solutions to address issues of privacy, confidentiality, and accessibility. Additionally, they explore emerging technologies such as blockchain and software-defined networks. In this context, a recent survey conducted by Imran Makhdoom and others comprehensively addresses security issues and threats posed by IoT devices [4]. Furthermore, they emphasized that the inherent safety provided by communication protocols does not sufficiently protect IoT devices from malicious malware and potentially endangering attacks. Hassan and his associates conducted a survey regarding security issues for IoT devices [1].

However, the authors solely focus on introducing solutions like authentication and lightweight encryption, not the issue of identifying IoT malware. Furthermore, Felt and colleagues [5] examined 46 pieces of mobile malware in the wild and collected a dataset to evaluate the effectiveness of mobile malware identification and prevention methods. Costin et al. [6] presented only a comprehensive review and analysis of all known classes of IoT malware without delving into approaches for identifying IoT malware. There are two main ways to find IoT malware: dynamic and static analysis. The dynamic way is to watch executable files while they run and find weird behaviors [7]. But this way is not very good, because some malware only acts badly when certain things happen. Also, it is hard to run IoT executable files because they use different architectures like MIPS, ARM, PowerPC, Sparc, and they have limited resources.

The static way is to look at and find bad files without running them. One big benefit of static analysis is that it can see what the malware is made of. This means that we can see all the possible ways the malware can work, no matter what kind of processor it uses. This is great for dealing with the variety of IoT devices.

However, there is not much research on how to use static analysis to find IoT malware, even though there are many studies on IoT security and malware detection. Identifying IoT malware is becoming a key issue for ensuring the security of the internet system and personal data. In summary, IoT malware identification methods can be divided into two groups: non-graph-based and graph-based methods. Non-graph-based methods can achieve good results when identifying malware without customization or obfuscation, but they potentially lose their accuracy when identifying clear malware. On the other hand, graph-based methods show advantages when analyzing the control flow of IoT malware, so despite the complexity of these methods, they have the potential to identify precise malicious codes that are invisible or complicated. Based on the mechanism, detection analysis and processing time, the advantages and limitations of the work done so far, they can be used to improve efficiency in future research. As a further development of this work, a lightweight graph-based detection method can be designed and developed that helps to detect malicious executable files in IoT devices. In this paper, the following solutions are presented to address the existing challenges:

> 1 - We receive the datasets and divide them into two groups for experimental and training sections. For this purpose, we use k-fold cross validation
>
> 2 - Since the data have a high volume, their dimensions must be reduced, so we use SVD. This method reduces linear dimensions using short singular value decomposition (SVD).
>
> 3 - In the next step, feature selection is performed. Therefore, we use Chi-Squared. This is another filter-based method. In this method, we calculate the chi-square metric between the target and the numerical variable and only select the variable with the maximum chi-square values and create a learning model with the help of CNN.

In the following, the background of the research will be discussed in the second section. The proposed method is explained in the third section. In the fourth section, the proposed method is simulated with the Python program and the results will be compared with other methods. In the fifth section, we will conclude and make suggestions for future work.

## 2. Background of the Research

The initial approach for identifying IoT devices on the internet involves scanning the entire 4Ipv address space. The use of scanners to find specific types of devices has been demonstrated, as shown in [8]. Researchers illustrate how Shodan, Masscan, and Nmap can be used to identify particular vulnerable IoT models. Similarly, CERN network researchers analyzed and attempted to identify IoT devices using web scraping [9]. They identified all devices, initially by scanning their open ports and subsequently by scraping their web interfaces when they had a port. Using this technique, they successfully identified numerous models and manufacturers of Internet of Things devices. They then conducted vulnerability assessments on the identified devices, noting that 11% were vulnerable by default and an additional 13% could be easily guessed through default credential verification.

The second intriguing approach to identify Internet of Things devices involves traffic capture. Although this is a new research area, some researchers have developed techniques to identify devices based on traffic capture. Most of these techniques are based on domain names that the devices contact, as they can be easily obtained from the captures as they are not encrypted in DNS lists. Authors in [10] propose a model for fingerprinting IoT devices behind NAT (Network Address Translation) and identifying them in a precise and explainable manner. Their idea involves indexing each device with a list of domains related to their query frequency. Since researchers cannot own all IoT devices, Princeton University researchers crowdsourced IoT identification and suggested Inspector-IoT, a tool aimed at collecting device traffic to create datasets for IoT identification. This tool is intended to be run by volunteer users who have devices on their personal computers.

Hang Guo and John Heidemann [11] attempted to identify Internet of Things traffic to measure the growth of IoT devices. They propose three detection techniques: IP-based, DNS-based, and TLS-based. The IP-based technique works by listening to DNS traffic to record IP address profiles of devices purchased. The second technique, DNS-based, is similar to the previous one, except that working with domain names instead of IP addresses prevents changes over time and facilitates identification of third-party and manufacturer domains thanks to domain names and WHOIS information. These techniques exhibit high accuracy in their own devices and seem to be more flexible over time as they have identified devices through several years of old traffic captures.

Finally, the TLS-based detection method aims to identify IoT devices that present an HTTPS interface (e.g., IP cameras). Researchers analyze web page TLS certificates and search for keywords that can identify the manufacturer or type of device. Previous studies primarily conducted traffic detection on a local scale, for example, within a university environment. Authors in [12] studied IoT device detection on a larger scale: transferring detection to the ISP (Internet Service Provider) or IXP (Internet Exchange Point) level. The detection technique relies on specific IoT infrastructure detection. Researchers initially identify domains that devices contact by monitoring DNS traffic and classify them into specific IoT domains and public domains. Then they obtain IP addresses associated with IoT-specific domains using DNSDB and filter out those shared across multiple services to derive dedicated infrastructure. Ultimately, the endpoint (IP address, port) is linked to the related device, creating a profile for each type of device. This method achieves excellent detection performance in evaluating functions; however, researchers acknowledge its limitations as it cannot identify devices that were not part of the training set and does not work well for devices with limited network traffic.

Malware analysis in the Internet of Things is performed using static, dynamic, and hybrid analysis techniques. Nazra and colleagues [13] were the first to perform malware analysis based on gray-scale images in 2011. Visual images of malware are created by rewriting the eight-bit code value of executable files to the corresponding gray-scale value. Texture features are extracted from these images. Texture-based analysis for Internet of Things malware in the field of deep learning is emerging. Evanson and colleagues [14] proposed an approach for analyzing malware using texture images of malware files and machine learning in IoTPOT for Bashlite and Mirai [15]. They provided Haralick texture features from the co-occurrence matrix and used machine learning classifiers.

Carillo and colleagues [16] examined the forensic and reverse engineering capabilities of malware for identifying IoT malware. They initially used machine learning to identify malware for Linux-based IoT systems. They also discovered new malware using clustering techniques. They utilized the dataset provided by E.Cozzi and colleagues [17].

Ganesh and colleagues [18] used machine learning capabilities to identify Mirai botnet attacks in the Internet of Things. They used ANN for evaluation in the BaIoT-N dataset. Bandiyab and colleagues [19] used deep learning for analyzing IoT traffic malware. They applied 50ResNet for empirical validation of their concept using a 1000-network file (pcap).

Kyushu and colleagues [20] proposed a lightweight approach to identify Internet of Things malware. They targeted DDoS malware for their study and extracted malware images from IoTPOT binaries. Their experimental setup demonstrated performance for identifying DDoS malware and benign software.

Ren and colleagues [21] presented a comprehensive malware detection mechanism for IoT Android devices. They collected 8000 malicious APK files and 8000 harmful files respectively from the Google Play store and VirusShare. They utilized deep learning importance for evaluating their concept.

Naeem and colleagues [22] identified industrial IoT malware with a proposed deep analysis of CNN-based traffic. They used color images of the intended malware for identification in the Mobile Leopard dataset.

## 3. The Proposed Method

In this section, the steps of the proposed method will be explained. This method consists of several phases, and we'll discuss each phase separately below:

**Phase One:**

In the first stage, the database is entered and divided into two parts: training and testing. We've used the UNSW-NB15 database in this context [23]. UNSW-NB15 is a dataset for network intrusion detection. This database includes 9 different attacks, such as DoS attacks, worms, backdoors, fuzzers, etc. It contains a number of raw network packets. This dataset is split into two sections: training set consisting of 175,341 records and a test set with 82,332 records, some of which are specific to attacks, while others are normal records. The data available in the UNSW-NB15 dataset was generated in the Cyber Range lab at UNSW Canberra using IXIA PerfectStorm tools. As mentioned earlier, it's formed from a combination of artificial attacks and normal events.

**Phase Two:**

In this stage, we examine the entered data and run the dimensionality reduction algorithm to reduce the problem's constraints and achieve better results. The SVD system is used for dimensionality reduction. SVD or Singular Value Decomposition is one of several techniques used to reduce the dimensions, i.e., the number of columns, in a dataset. In predictive analytics, more columns usually mean more time to build models and score data. If some columns do not contribute to prediction value, it's a waste of time, or worse, they add noise to the model and reduce its quality or predictive accuracy.

Dimensionality reduction can simply be achieved by discarding columns, for example, those that may be linearly dependent or specifically non-predictive, identified by feature importance ranking techniques. However, new columns based on linear combinations of the original columns can also be extracted. In either case, the transformed dataset obtained can be fed into machine learning algorithms to achieve faster model building, quicker scoring time, and more accurate models. While SVD can be used for dimensionality reduction, it's often used in digital signal processing for noise reduction, image compression, and other domains.

SVD [26] is an algorithm that takes an m x n matrix, M, of real or complex values and decomposes it into three component matrices, represented by USV*. U is an m x p matrix. S is a diagonal matrix of size p x p. V is an n x p matrix, with V* as the transpose of V, a p x n matrix. If M contains complex values, p is referred to as the rank. The diagonal entries of the S matrix are known as the singular values of M. The columns of U are usually called the left singular vectors of M, and the columns of V are the right singular vectors of M.

Consider the visual representation of these matrices below:

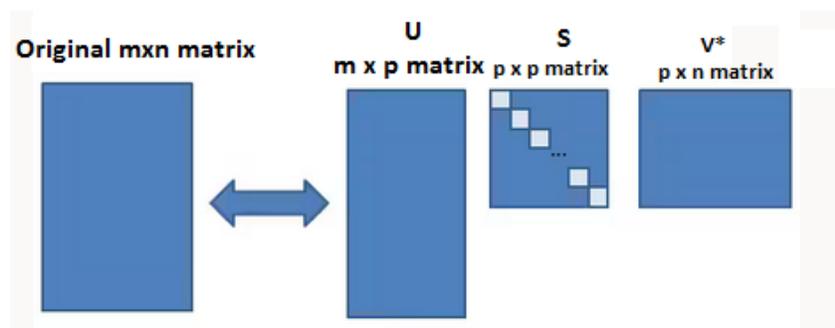

Figure 1 - Data Matrix Components in SVD (Rokhlin et al., 2019)

One of the features of SVD is that, based on the decomposition of M into U, S, and V, the original matrix M or an approximation of it can be reconstructed. The singular values in the diagonal matrix S can be used to understand the amount of variance explained by each of the singular vectors. Figure 2 shows the pseudo-code algorithm for SVD.

**Algorithm 1** SVD

**Require:** $A$ (real $m \times n$ matrix, without loss of generality assume $m \leq n$), $k$ (desired rank of truncated SVD), $p$ (parameter for oversampling dimension), $\ell = k + p < m$ (dimension of the approximate column space), $q$ (exponent of the power method)

**Ensure:** Approximate rank-$k$ SVD of $A \approx \widehat{U}_k \widehat{\Sigma}_k \widehat{V}_k^\top$

1: Generate an $n \times \ell$ random matrix $\Omega$.
2: Assign $Y \leftarrow (AA^\top)^q A\Omega$.
3: Compute $Q$ whose columns are an orthonormal basis of $Y$.
4: Compute the SVD of $Q^\top A = \widehat{W}_\ell \widehat{\Sigma}_\ell \widehat{V}_\ell^\top$.
5: Assign $\widehat{U}_\ell \leftarrow Q\widehat{W}_\ell$.
6: Extract the leading $k$ singular vectors and singular values from $\widehat{U}_\ell, \widehat{\Sigma}_\ell$ and $\widehat{V}_\ell$ to obtain $\widehat{U}_k, \widehat{\Sigma}_k$ and $\widehat{V}_k$.

Figure 2 - Pseudo-code of the SVD Dimension Reduction Method (Anowar, 2021)

**Phase 3:**

Following dimension reduction, the feature selection stage begins. Feature selection methods are commonly categorized into three classes: 1 - filter-based methods, 2 - wrapper-based methods, and 3 - embedded methods, with the best being machine-based methods. In this study, the Convolutional Neural Network (CNN) method has been employed for feature selection.

Feature Selection Algorithm:

1 - Let A = {A1, A2, …An} be a set of input features to the CNN. Assume R is the maximum acceptable drop in accuracy of the test set.

2 - Train network N to minimize loss values with input A until the accuracy of the training set is acceptable.

3 - For all k = 1, 2, …n, set the weights of input Ak to zero in network Nk while keeping the weights of other inputs equal to the weight of network N.

4 - Calculate the accuracy of the training set (Rk) and the test set (R'k) respectively.

5 - Rank the Nk networks based on the accuracy of the training set.

6 - Compute the change in test set accuracy r for each Nk starting from k = 1. If r <= R, remove Ak from the input set A, and N = N-1. If k < N, k = k+1, and proceed to the next iteration. Otherwise, stop the algorithm.

**Phase 4:**

After model training, batch operations need to be performed. Among the top 10 styles introduced in 2022, the LSTM system is used for data classification. LSTMs are a type of Recurrent Neural Network (RNN) capable of learning and memorizing long-term dependencies. LSTMs work in a three-stage process.

Stage 1: Decide how much past data should be remembered.

The first stage in LSTM is deciding which information should be discarded at a particular time step from the cell. The sigmoid function determines this. It looks at the previous state (ht-1) along with the current input (xt) and computes the function. In LSTM [27], the inference for cell states (ct) and hidden states (ht) is represented by the following equation (Biswal, 2023):

$$c_t = f_t \odot c_{t-i} + i_t \odot \tanh(U h_{t-1} + W x_t + b) \quad (1)$$
$$h_t = o_t \odot \tanh(c_t) \quad (2)$$

The symbol ⊙ represents an element-wise multiplication operation, and within this framework, three gates are defined as follows according to Biswal (2023):

$$i_t = \sigma(W_i x_t + U_i h_{t-1} + b_i) \quad (3)$$
$$f_t = \sigma(W_f x_t + U_f h_{t-1} + b_f) \quad (4)$$
$$o_t = \sigma(W_o x_t + U_o h_{t-1} + b_o) \quad (5)$$

Here, σ represents the sigmoid function allowing the input information to influence the output in the current time step. $i_t$ determines which information to release based on its importance at the current time step. $f_t$ decides which information can be discarded as it's not crucial from the previous information, where U, W, and b are input parameters.

## 4. Results and Evaluation

As simulation and evaluation stand among the most crucial segments of any research work, this section delves into evaluating the proposed method. In the preceding section, a solution for intrusion detection in a network was presented. This part focuses on simulation outcomes and their comparison based on multiple evaluation metrics. Python software is employed for simulations.

**4.1 Required Datasets**

To evaluate the proposed method, simulations must be conducted based on certain existing datasets, and the performance of the proposed method must be evaluated. In this regard, the following datasets have been used in this study:

### 4.1.1 - BoT-IoT

The BoT-IoT dataset was crafted by designing a real network environment in the Cyber Range laboratory at UNSW Canberra. This network environment encapsulates a blend of regular traffic and botnet behavior. The dataset source files are provided in various formats, including primary pcap files, generated argus files, and csv files. The files have been segregated based on attack categories and subcategories to facilitate better labeling in the labeling process.

The pcap files collected amount to 69.3 gigabytes with over 72,000,000 records. The extracted traffic in CSV format stands at 16.7 gigabytes. The dataset encompasses DDoS, DoS, OS and Service Scan, Keylogging, and Exfiltration Data attacks, predominantly organized based on the protocol used.

### 4.1.2 - CSE CIC IDS2018

The CSE CIC IDS2018 dataset is tailored for training models capable of identifying or predicting network intrusions. This dataset, along with similar ones, is employed in research aiming to detect network intrusions using machine learning algorithms.

The CSE CIC IDS2018 dataset comprises nearly 160,000,000 data samples. This dataset stands as the most recent intrusion detection dataset, allowing access to substantial data volumes and encompassing a wide array of network attack types.

### 4.2 Evaluation Methods

**False Positive Rate (FPR) Metric:** It examines what percentage of intruded states are falsely identified as normal, giving insight into the system's misidentification percentage. Lower values indicate better performance, calculated using the formula $FPR=FP/N$, where FP represents falsely identified positive data, and N is the total count of abnormal vectors [24].

$$FPR = \frac{FP}{N} \quad (6)$$

**False Rejection Rate (FRR) Metric:** It demonstrates how often the intrusion detection system incorrectly marks authentic and correct states as intrusions, resulting in false alarms that restrict authorized users' network access. Calculated as $FRR=FN P$, where FN represents falsely identified negative data, and P is the total count of positive data [24].

$$FRR = \frac{FN}{P} \quad (7)$$

**Accuracy Metric:** This metric illustrates the system's percentage of correctly identifying intrusion and non-intrusion states. Calculated using the formula $ACC=true\ result all\ record$, higher values denote better performance [24].

$$ACC = \frac{true\ result}{all\ record} \quad (8)$$

**Precision Metric:** An important measure in intrusion detection algorithms, Precision, is calculated using $Precision=TP/TP+FP$, demonstrating the system's ability to correctly identify positive cases among all identified cases. Higher precision indicates fewer false positives [24].

$$Precision = \frac{TP}{TP+FP} \quad (9)$$

In this context, TP represents the number of data correctly identified as positive, while FP signifies the count of data mistakenly labeled as positive.

**Recall Metric:** It is calculated using the below formula [24]:

$$Recall = \frac{TP}{TP + FN} \quad (10)$$

Here, TP refers to the count of data correctly identified as positive, while FN represents the count of data incorrectly labeled as negative.

**F-measure Metric:** This metric, computed using formula (11), acts as a balance between recall and precision [24]:

$$Recall = \frac{TP}{TP + FN} \quad (11)$$

### 4.3 Dataset Results

```python
In [ ]: def LSTM_Network(neurons=100):
            model = tf.keras.models.Sequential()
            .
            .
            model.compile(loss='binary_crossentropy',
                    optimizer=RMSopt,
                    metrics=['accuracy'])
            return model

        lstm_clf = KerasClassifier(build_fn=LSTM_Network, epochs=6, batch_size=64, verbose=0)
        model_selection.cross_val_score(lstm_clf, X_train, Y_test, cv=10, scoring='accuracy')
```

{'test_accuracy': array([0.956987, 0.953247, 0.9535432, 0.954535532 , 0.951323]), 'test_precision': array([0.96587425, 0.9687256, 0.9683247, 0.968969 , 0.967412]), 'test_recall': array([0.95919268, 0.96018529, 0.96073241, 0.95819546, 0.96007059]), 'test_f1': array([0.94875369, 0.9485234, 0.947852, 0.9436589, 0.9459874]), 'test_roc_auc': array([0.9658245 , 0.9325874, 0.9758236, 0.9845632, 0.998574252])} 0.9425869325478

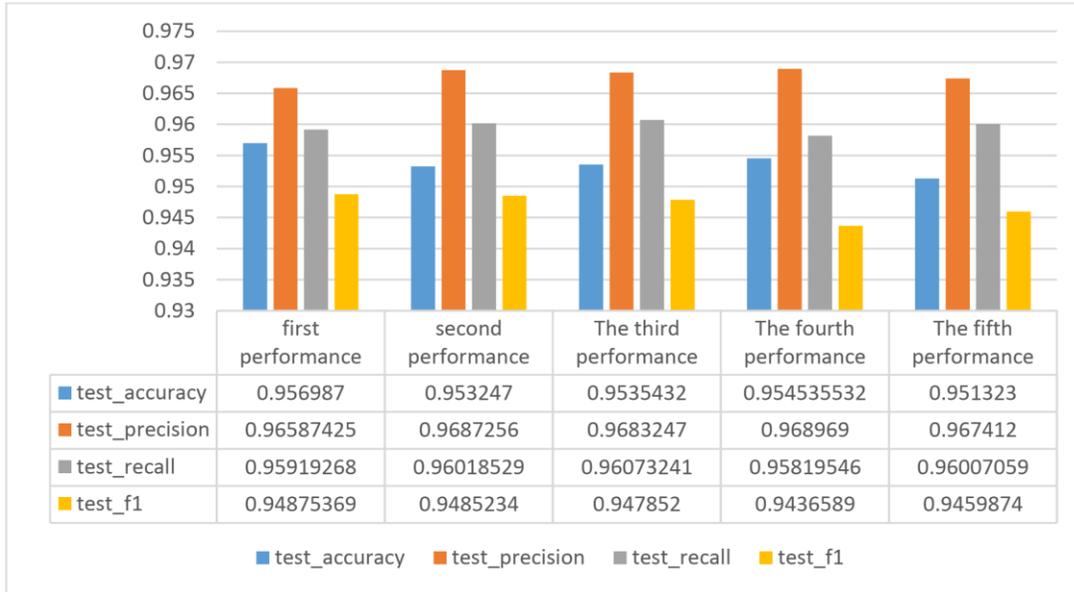

Chart -1: A comparison chart of results in the dataset

As shown in Chart 1, the proposed method was executed five times on two datasets, measuring its performance and detection capability each time based on precision, accuracy, recall, and the F-measure values. Due to different subsets being selected as the training set and models being built based on them, diverse results were obtained. To address this, multiple runs were conducted to display the final outcome as an average on the final charts and compare it with other works [19], [21].

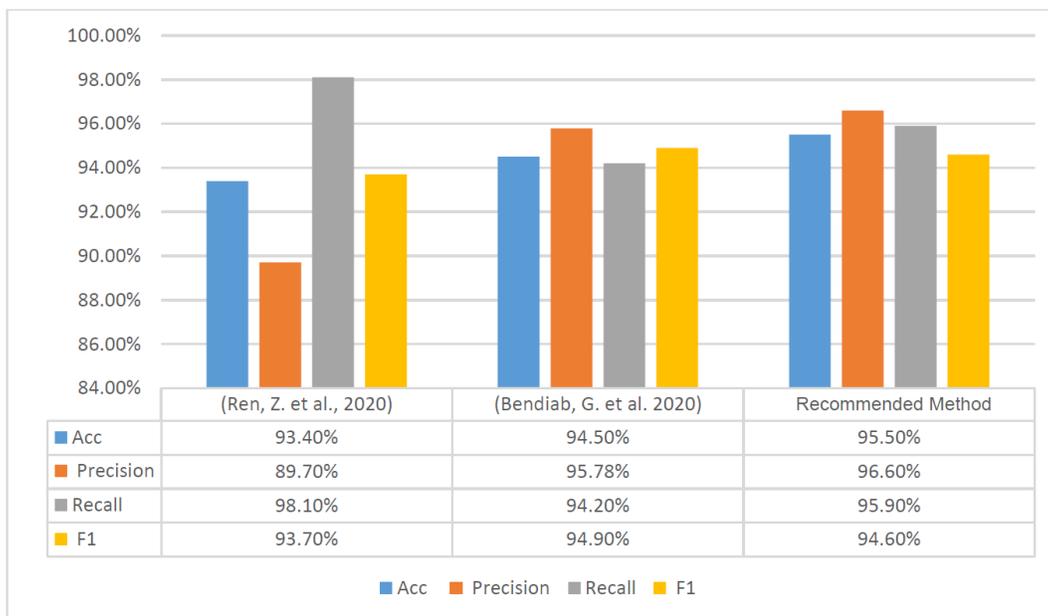

Chart 2 - Comparison of average results of the proposed method with other methods [19],[21]

Considering the results, the proposed method has demonstrated better capabilities in data analysis and intrusion detection within a shorter time frame. Moreover, compared to methods [21] and [19], it exhibits higher accuracy in intrusion detection. Method [21] achieved an accuracy of 93.4%, method [19] attained 94.5%, while the proposed method managed to detect network intrusions with an accuracy of 95.5%.

## 5. Discussion and Conclusion:

Effective malware detection demands appropriate and robust methods across diverse environments. Deploying autonomous systems is critical for network security, yet strategies like centralized or distributed detection pose performance trade-offs. Our study showcased a CNN-LSTM approach achieving 95.5% accuracy in IoT malware detection, surpassing existing methods [21] and [19]. This emphasizes the potential of this approach while signaling promising pathways for future research. We propose exploring SVMs as alternatives [25] and advocating for advanced techniques to fortify IoT security. Future endeavors should prioritize distributed detection methods to reduce performance bottlenecks and measure reliability against severe threats. Incorporating predictive analyses and deep learning model evaluations will further enhance detection systems. This study underscores the urgency for advanced strategies in IoT security, highlighting the need for tailored detection mechanisms and architectural theories to strengthen network defenses. In conclusion, our research contributes a robust methodology for IoT malware detection, but further investigations, enhancing distributed detection strategies, and integrating predictive analyses, are vital to fortify network security against evolving threats. This study sets the stage for future research directions aiming to create more resilient and adaptive security measures for IoT ecosystems.

## References:

[1] Wan Haslina, H., 2019. Current research on Internet of Things (IoT) security: A survey.
[2] Görmüş, S., Aydın, H. & Ulutaş, G., 2018. Security for the internet of things: a survey of existing mechanisms, protocols and open research issues.
[3] Kouicem, D., Bouabdallah, A. & Lakhlef, H., 2018. Internet of things security: A top-down survey.
[4] Makhdoom, I. et al., 2018. Anatomy of threats to the internet of things.
[5] Felt, A. et al., 2011. A survey of mobile malware in the wild.
[6] Costin, A. & Zaddach, J., 2018. IoT malware: Comprehensive survey, analysis framework and case studies.
[7] Nguyen, H., Ngo, Q. & Le, V., 2020. A novel graph-based approach for IoT botnet detection.
[8] Markowsky, L. & Markowsky, G., 2015. Scanning for vulnerable devices in the Internet of Things.
[9] Agarwal, S., Oser, P. & Lueders, S., 2019. Detecting IoT devices and how they put large heterogeneous networks at security risk.
[10] Perdisci, R., Papastergiou, T., Alrawi, O. & Antonakakis, M., 2020. IoTfinder: Efficient large-scale identification of iot devices via passive dns traffic analysis.
[11] Guo, H. & Heidemann, J., 2020. Detecting iot devices in the internet.
[12] Saidi, S. et al., 2020. A haystack full of needles: Scalable detection of IoT devices in the wild.
[13] Karthikeyan, L., Jacob, G. & Manjunath, B., 2011. Malware images: Visualization and automatic classification.

[14] Karanja, E., Masupe, S. & Jeffrey, M., 2020. Analysis of internet of things malware using image texture features and machine learning techniques.
[15] Pa, Y. et al., 2015. IoTPOT: Analysing the Rise of IoT Compromises.
[16] Carrillo-Mondéjar, J., Martínez, J. & Suarez-Tangil, G., 2020. Characterizing Linux-based malware: Findings and recent trends.
[17] Cozzi, E., Graziano, M., Fratantonio, Y. & Balzarotti, D., 2018. Understanding Linux malware.
[18] Palla, T. & Tayeb, S., 2021. Intelligent Mirai Malware Detection in IoT Devices.
[19] Bendiab, G., Shiaeles, S., Alruban, A. & Kolokotronis, N., 2020. IoT malware network traffic classification using visual representation and deep learning.
[20] Su, J. et al., 2018. Lightweight classification of IoT malware based on image recognition.
[21] Ren, Z. et al., 2020. End-to-end malware detection for android IoT devices using deep learning.
[22] Naeem, H. et al., 2020. Malware detection in industrial internet of things based on hybrid image visualization and deep learning model.
[23] Nour Moustafa, 2019. UNSW-NB15: A Comprehensive Data set for Network Intrusion Detection systems (UNSW-NB15 Network Data Set)
[24] Ngo, Q.-D., Nguyen, H.-T., Nguyen, L.-C. & Nguyen, D.-H., 2020. A survey of IoT malware and detection methods based on static features.
[25] P Ahadian, K Parand, 2022. Support vector regression for the temperature-stimulated drug release, Chaos, Solitons & Fractals 165 (Part 2), 112871
[26] Ahadian, Pegah, Maryam Babaei, and Kourosh Parand. "Using singular value decomposition in a convolutional neural network to improve brain tumor segmentation accuracy." arXiv preprint arXiv:2401.02537 (2024).
[27] Mehrban, Ali, and Pegah Ahadian. "evaluating bert and parsbert for analyzing persian advertisement data." arXiv preprint arXiv:2305.02426 (2023).